\documentclass[aps,prl,twocolumn,superscriptaddress]{revtex4-1}
\usepackage{amssymb, amsmath}
\usepackage{graphicx,onlyamsmath}

\usepackage{natbib}
\usepackage{xcolor}

\begin{document}

%\title{Magneto-optical study of InAs/GaSb three-layers}
\title{Large gap quantum spin Hall insulator, massless Dirac fermions and bilayer graphene analogue in InAs/Ga(In)Sb heterostructures}

\author{S.~S.~Krishtopenko}
\affiliation{Laboratoire Charles Coulomb, UMR Centre National de la Recherche Scientifique 5221, Universit\'{e} de Montpellier, 34095 Montpellier, France.}
\affiliation{Institute for Physics of Microstructures RAS, GSP-105, 603950, Nizhni Novgorod, Russia.}

\author{F.~Teppe}
\email[]{frederic.teppe@umontpellier.fr}
\affiliation{Laboratoire Charles Coulomb, UMR Centre National de la Recherche Scientifique 5221, Universit\'{e} de Montpellier, 34095 Montpellier, France.}
\date{\today}% It is always \today, today,

\begin{abstract}
The quantum spin Hall insulator (QSHI) state has been demonstrated in two semiconductor systems -- HgTe/CdTe quantum wells (QWs) and InAs/GaSb QW bilayers. Unlike the HgTe/CdTe QWs, the inverted band gap in InAs/GaSb QW bilayers does not open at the $\Gamma$ point of the Brillouin zone, preventing the realization of massless Dirac fermions. Here, we propose a new class of semiconductor systems based on InAs/Ga(In)Sb multilayers, hosting a QSHI state, a graphene-like phase and a bilayer graphene analogue, depending on their layer thicknesses and geometry. The QSHI gap in the novel structures can reach up to 60 meV for realistic design and parameters. This value is twice as high as the thermal energy at room temperature and significantly extends the application potential of III-V semiconductor-based topological devices.
\end{abstract}

%Use showkeys class option if keyword                            %display desired
\maketitle
\emph{Introduction.}\textbf{--}Quantum spin Hall insulators (QSHIs), also known as two-dimensional (2D) topological insulators (TIs), are characterized by an insulating bulk and spin-polarized gapless helical states at the sample edges~\cite{Q1,Q2,Q3}. Those edge states are protected from backscattering by time reversal symmetry, promoting dissipationless electric currents. QSHIs also offer realizations of numerous non-trivial properties, such as unusual magnetoelectric effects~\cite{Q4} and Majorana fermion bound states, arising at the interface with superconductors~\cite{Q5,Q6}. Experimental demonstrations of QSHI are so far limited to two semiconductor systems -- HgTe/CdTe quantum wells (QWs)~\cite{Q7} and InAs/GaSb bilayers~\cite{Q8,Q9}.

The central feature of the HgTe/CdTe QWs is the possibility of band inversion by changing of QW width $d$. QWs wider than a critical thickness $d_c$ have inverted band structure~\cite{Q10}, at which the lowest electron-like subband (\emph{E}1) lies lower in energy than the top heavy-hole-like subband (\emph{H}1). The band inversion drives the system into the QSHI state~\cite{Q3}, with gapless edge channels. At $d<d_c$, the \emph{E}1 subband lies above \emph{H}1 level and the systems features the band insulator (BI) state. At the critical thickness, the HgTe QWs possess a single-valley spin-degenerate Dirac cone at the $\Gamma$ point of the Brillouin zone~\cite{Q11}. The maximum inverted band gap in HgTe/CdTe QWs grown on CdTe does not exceed 16 meV. Recently, P. Leubner~\emph{et~al.} have reported realization of QSHI with a band gap up to 55 meV in compressively strained HgTe QWs~\cite{Q12}.

The InAs/GaSb bilayers consist of two layers of InAs and GaSb confined between AlSb barriers. The valence band edge of GaSb is 0.15 eV higher than the conduction band edge of the InAs layer, thus, depending on the layer thicknesses, a band inversion arises. Since the InAs/GaSb bilayers do not have inversion symmetry in the growth direction, the crossing of \emph{E}1 and \emph{H}1 subbands arises away from the $\Gamma$ point of the Brillouin zone~\cite{Q13}. Therefore, this band gap vanishing does not reveal a Dirac cone due to the mixing between \emph{E}1 and \emph{H}1 levels at non-zero quasimomentum $k$, which opens a hybridization gap and drives the system into the QSHI regime. The typical band gap values in the inverted InAs/GaSb bilayers are of several meV~\cite{Q8}.

%Note that replacing of GaSb by GaInSb alloy in the QW bilayers enhances to the gap up to 20 meV for realistic indium content~\cite{Q14,Q15}.

\begin{figure}
\includegraphics [width=1.0\columnwidth, keepaspectratio] {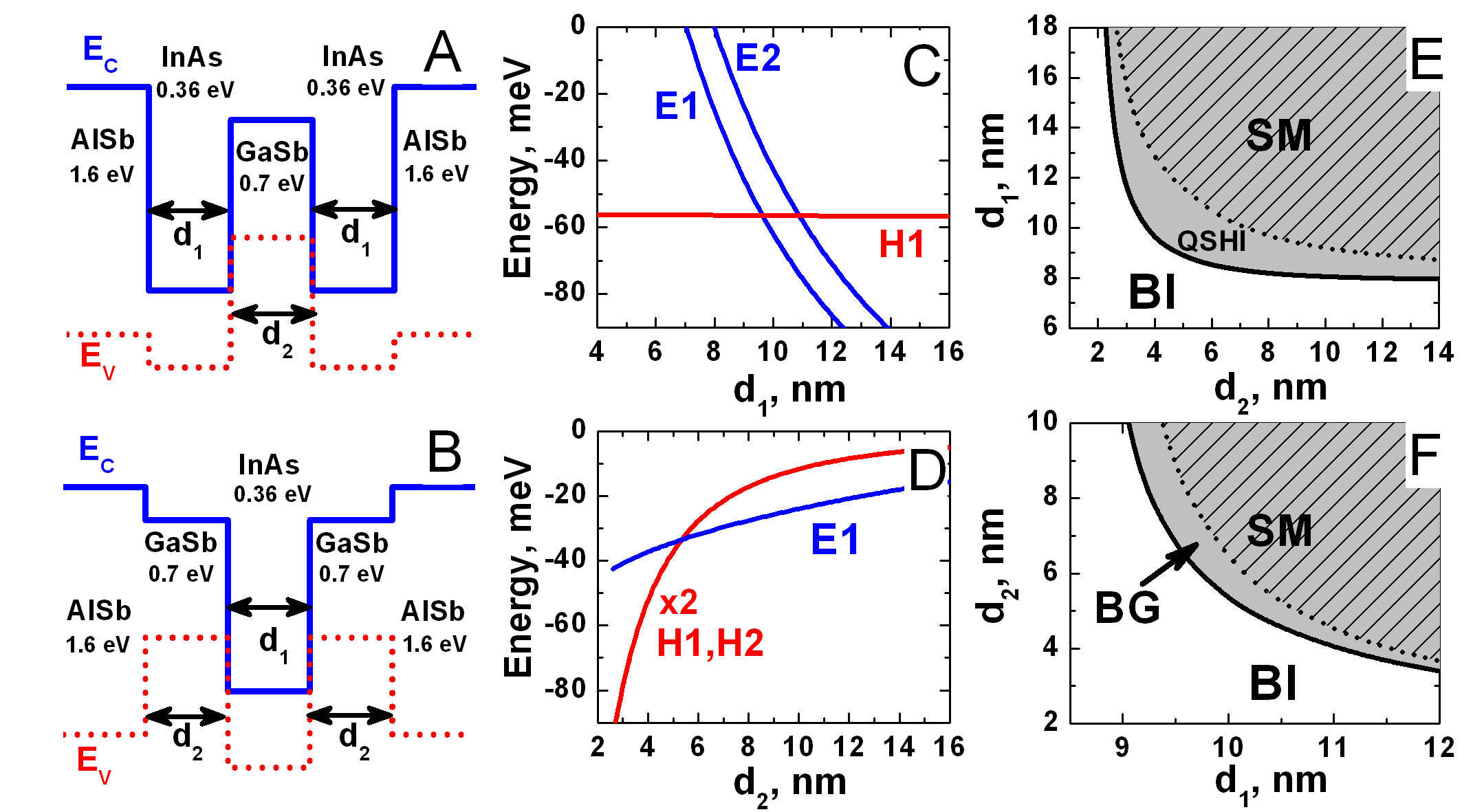} % Here is how to import EPS art
\caption{\label{Fig:1} (A,B) Schematic representation of symmetrical three-layer InAs/GaSb QWs: (A) InAs-designed QWs and (B) GaSb-designed QWs. The numbers show the band gap values in materials of the layers. (C,D) Energy of electron-like (blue curves) and heavy-hole like (red curves) subbands at $k=0$, as a function of the layer thickness: (C) InAs-designed QW at $d_2 = 4$~nm and (D) GaSb-designed QW at $d_1=10$~nm. Crossing of \emph{E}1 and \emph{H}1 levels produces massless Dirac fermions in the vicinity of the $\Gamma$ point of the Brillouin zone. Zero-energy corresponds to the top of valence band in bulk GaSb. (E,F) Phase diagram for different $d_1$ and $d_2$ in (E) InAs-designed QWs and (F) GaSb-designed QWs.}
\end{figure}

In this work, we introduce a new class of multi-layered InAs/Ga(In)Sb QWs, which differ from the bilayers by the band crossing at the à point of the Brillouin zone. To eliminate the inversion asymmetry in the growth direction, we attach an additional InAs or GaSb layer to the InAs/GaSb bilayer. Further, we call the three-layer InAs/GaSb QW as InAs-designed QW (Fig.~\ref{Fig:1}A) or GaSb-designed QW (Fig.~\ref{Fig:1}B) if it contains two layers of InAs or GaSb, respectively.

\emph{Phase transitions in three-layer InAs/GaSb QWs.}\textbf{--} \\ To investigate the band ordering, we have performed band structure calculations on the basis of the full 8-band Kane model~\cite{Q16} with material parameters taken from~\cite{Q17}. Figure~\ref{Fig:1}C shows positions of electron-like and hole-like subbands at zero quasimomentum $k$ in the InAs-designed QW as a function of InAs-layer thickness $d_1$. The QWs are supposed to be grown on GaSb buffer along (001) crystallographic direction.

As the electron-like subbands are localized in the InAs layers, while the hole-like subbands are localized in the GaSb layers, the InAs-designed QW can be contingently considered as "double QW for electrons" with GaSb middle barrier ($d_2=4$~nm), which also plays a role of "QW for holes". In this case, the electron-like levels \emph{E}1 and \emph{E}2 are connected with even-odd state splitting, arising from the tunnel-coupled "QWs for electrons". If InAs layers are thin ($d_1<d_{1c}$, $d_{1c}$ is a function of $d_2$), the \emph{E}1 and \emph{E}2 subband lie above the position of valence band edge in bulk GaSb, and the three-layer QW is in normal regime. For $d_1>d_{1c}$, the \emph{E}1 subband lies below the \emph{H}1 level and the QW has an inverted band ordering.

In contrast, the GaSb-designed QW contains two tunnel-coupled "QWs for holes" with InAs layer as "middle barrier". The InAs layer at $d_1>2$~nm is not transparent for the particles from the hole-like subbands due to their large effective mass. Therefore, the hole-like subbands are degenerated at $k=0$. The mixing between heavy-hole-like and electron-like states at non-zero $k$ leads to the splitting between \emph{H}1 and \emph{H}2 subbands. Figure~\ref{Fig:1}D shows positions of \emph{E}1, \emph{H}1 and \emph{H}2 subbands at $k=0$ in GaSb-designed QW with $d_1=10$~nm as a function of $d_2$. If GaSb layers are thin enough so that $d_2<d_{2c}$ (where $d_{2c}$ depends on $d_1$), the \emph{E}1 level lies below \emph{H}1 and \emph{H}2 subbands, and the band structure is inverted. For $d_2>d_{2c}$, the QW has direct band structure.

For both types of three-layer structures, we have explored the phase diagrams, shown in Fig.~\ref{Fig:1}E and \ref{Fig:1}F. In each diagram, the solid curve, describing the crossing between \emph{E}1 and \emph{H}1 subbands, divides the $d_1$--$d_2$ plane into white region, corresponding to the BI phase with direct band ordering, and grey region with inverted band structure.

The grey region for InAs-designed QWs is splitted into open and striped areas, corresponding to QSHI and semimetal (SM) phase, respectively. The SM phase is characterized by a vanishing indirect band gap, when the side maxima of the valence subband exceed in energy the conduction subband bottom. This phase is also present in wide HgTe QWs~\cite{Q18,Q19}. The typical band dispersions for each phase are presented in~\cite{Q20}.

The band inversion in GaSb-designed QWs drives the BI state into a metal phase with a band structure consisting of two isotropic parabolas, formed by conduction subband H1 and valence subband H2 touching at $k=0$. Such band structure is very similar to the one of natural bilayer graphene (BG)~\cite{Q21}. The reasons for naming this metal phase as a BG phase will be discussed later. Further increasing of $d_1$ and $d_2$ transforms a BG phase into a metal phase, which, in addition to the band touching at $k=0$, is also characterized by non-local overlapping of the valence subband with the loop of side minima in the conduction subband. This unconventional SM phase is presented by grey-striped area in Fig.~\ref{Fig:1}F. Typical band dispersion for each phase is provided in~\cite{Q20}.

To describe quantum phase transition, we derive effective 2D Hamiltonian for both types of the QWs. In addition to \emph{E}1 and \emph{H}1 levels, we should also take into account adjacent \emph{E}2 subband for the InAs-designed QWs and \emph{H}2 subband for the GaSb-designed QWs. Following the procedure described in~\cite{Q3,Q22}, 2D Hamiltonian for the InAs-designed QWs in the basis $|$\emph{E}1,+$\rangle$, $|$\emph{H}1,+$\rangle$, $|$\emph{E}2,-$\rangle$, $|$\emph{E}1,-$\rangle$, $|$\emph{H}1,-$\rangle$, $|$\emph{E}2,+$\rangle$ has the form:
\begin{equation}
\label{eq:1}
H_{2D}(k_x,k_y)=\begin{pmatrix}
H_1(k_x,k_y) & 0 \\ 0 & H_1^{*}(-k_x,-k_y)\end{pmatrix},
\end{equation}
where asterisk stands for complex conjugation, $k_x$ and $k_y$ are momentum $k$ components in the plane, $k_{\pm}=k_x\pm ik_y$ and $H_1(k_x,k_y)$ is written as
\begin{equation}
\label{eq:2}
H_1^{(InAs)}(k_x,k_y)=\begin{pmatrix}
\varepsilon_{E1}(k) & -Ak_{+} & Sk_{-} \\ -Ak_{-} & \varepsilon_{H1}(k) & Rk_{-}^2 \\ Sk_{+} & Rk_{+}^2 & \varepsilon_{E2}(k)\end{pmatrix},
\end{equation}
in which $\varepsilon_{E1}(k)=C+M+B_{E1}k^2$, $\varepsilon_{H1}(k)=C-M+B_{H1}k^2$ and $\varepsilon_{E2}(k)=C+M+\Delta_{E1E2}+B_{E2}k^2$. Here, $C$, $M$, $A$, $B_{E1}$, $B_{H1}$, $B_{E2}$, $R$, $S$ are structure parameters, which depend on $d_1$ and $d_2$ and $k^2=k_{+}k_{-}$. The sign of mass parameter $M$ defines inversion of the \emph{E}1 and \emph{H}1 levels. Additional parameter $\Delta_{E1E2}$ describes the gap between the \emph{E}1 and \emph{E}2 subbands at $k=0$. Since we have kept the inversion symmetry and axial symmetry around the growth direction, $H_{2D}(k_x,k_y)$ has a block-diagonal form.

In the basis $|$\emph{E}1,+$\rangle$, $|$\emph{H}1,+$\rangle$, $|$\emph{H}2,-$\rangle$, $|$\emph{E}1,-$\rangle$, $|$\emph{H}1,-$\rangle$, $|$\emph{H}2,+$\rangle$, 2D Hamiltonian for the GaSb-designed QWs has the form of (\ref{eq:1}) but with $H_1(k_x,k_y)$ defined as
\begin{equation}
\label{eq:4}
H_1^{(GaSb)}(k_x,k_y)=\begin{pmatrix}
\varepsilon_{E1}(k) & -Ak_{+} & Rk_{-}^2 \\ -Ak_{-} & \varepsilon_{H1}(k) & 0 \\ Rk_{+}^2 & 0 & \varepsilon_{H2}(k)\end{pmatrix},
\end{equation}
where $\varepsilon_{H2}(k)=C-M+B_{H2}k^2$. Here, the terms $\varepsilon_{E1}(k)$ and $\varepsilon_{H1}(k)$ have the same form as in Eq.~(\ref{eq:2}), $B_{H2}$ depends only on thickness of InAs and GaSb layers. Parameters for both 2D Hamiltonians and comparison with band structure calculations on the basic of the 8-band Kane model are given in~\cite{Q20}. At $M=0$, both Hamiltonians are reduced to Hamiltonian of massless Dirac fermions with Fermi velocity $v_F$, defined by parameter $A$. The values of $v_F$ in three-layer InAs/GaSb QWs are varied from $1\cdot10^5$~m/s to $3\cdot10^5$~m/s depending on the layer thicknesses and buffer material~\cite{Q20}.

\begin{figure}
\includegraphics [width=1.0\columnwidth, keepaspectratio] {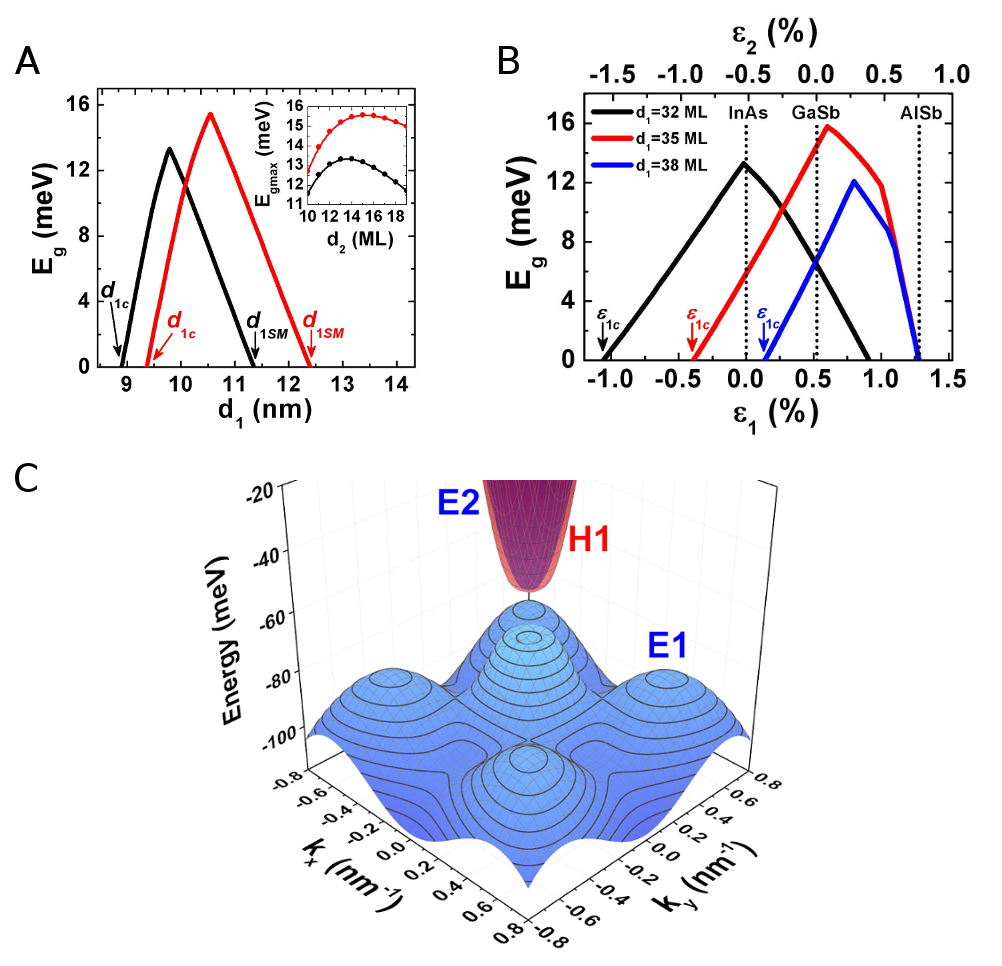} % Here is how to import EPS art
\caption{\label{Fig:2} (A) Band gap in QSHI state as a function of $d_1$ in InAs/GaSb/InAs QW with $d_2=14$~ML. The black and red curves correspond to the structures grown on InAs and GaSb buffer, respectively. The inset shows the maximum gap, which can be achieved at given value of $d_2$. (B) Inverted band gap as a function of $\varepsilon_1$ and $\varepsilon_2$. (C) A 3D plot of the band dispersion for InAs-designed QW with $d_1=35$~ML and $d_2=14$~ML grown on GaSb buffer. The $x$ and $y$ axes are oriented along (100) and (010) crystallographic directions, respectively.}
\end{figure}

\emph{QSHI in InAs-designed QWs.}\textbf{--}QSHI state in the InAs-designed QWs occurs in the inverted regime, in which the massless helical states are confined on the sample edge. The latter can be shown by solving the eigenvalue problem for $H_1^{(InAs)}(k_x,k_y)$ with $M<0$ and open boundary condition along one of the directions~\cite{Q20}.

In the bilayers, the inverted gap opens away from the $\Gamma$ point of the Brillouin zone due to the mixing of \emph{E}1 and \emph{H}1 levels. In the case of three-layer InAs/GaSb QWs, the inverted band gap arises at $k=0$ due to quantum confinement, while the \emph{E}1-\emph{H}1 mixing vanishes at $k=0$. Since the confinement effect is much stronger then the \emph{E}1-\emph{H}1 interaction at non-zero $k$, we expect higher values of the inverted band gap in the novel structures than in InAs/GaSb bilayers.

Figure~\ref{Fig:2}A shows the band gap in QSHI state as a function of $d_1$ for $d_2=14$ monolayers (ML). Here, 1 ML in the given QW layer corresponds to a half of a lattice constant $a_0$ in the bulk material. It is seen that, at $d_1>d_{1c}$, increasing of $d_1$ induces increasing of the inverted band gap, which occurs between \emph{H}1 and \emph{E}1 levels. When the band gap becomes close to $\Delta_{E1E2}$ (see Fig.~\ref{Fig:2}C), the system transforms into indirect-band-gap QSHI, in which \emph{H}1 level lies above \emph{E}2 subband. Further increasing of $d_1$ decreases the indirect-band-gap between $E_1$ and $E_2$ subbands until it vanishes at $d_1=d_{1SM}$. The latter causes a transition into SM phase. The values of $d_{1c}$ and $d_{1SM}$ also depend on the strain in InAs and GaSb layers, which can be varied by changing the buffer material.

Figure~\ref{Fig:2}B provides the inverted band gap as a function of strain in InAs layers $\varepsilon_1$ and GaSb layer $\varepsilon_2$. The values of $\varepsilon_1$ and $\varepsilon_2$ are connected via the lattice constant of the buffer material. It is seen that by tuning the strain one can induce transitions between BI, QSHI and SM phases. The value of $\varepsilon_1$, corresponding to the crossing between \emph{E}1 and \emph{H}1 level, is marked by vertical arrow and defined as $\varepsilon_{1c}$. It is seen that by adjusting the strain and the values of $d_1$ and $d_2$, one can realized a QSHI with a band gap up to 16 meV. For instance, the QW with $d_1=35$~ML and $d_2=14$~ML grown on GaSb buffer has a QSHI state with a gap close to the maximum value (see Fig.~\ref{Fig:2}C). The band gap of 16 meV is in several times greater than the band gap in inverted InAs/GaSb bilayers~\cite{Q8}. It is interesting that the QWs grown on AlSb buffer do not feature a QSHI state. In this case, the band inversion at $k=0$ is accompanied by nonlocal overlapping between conduction and valence bands.

\begin{figure}
\includegraphics [width=0.9\columnwidth, keepaspectratio] {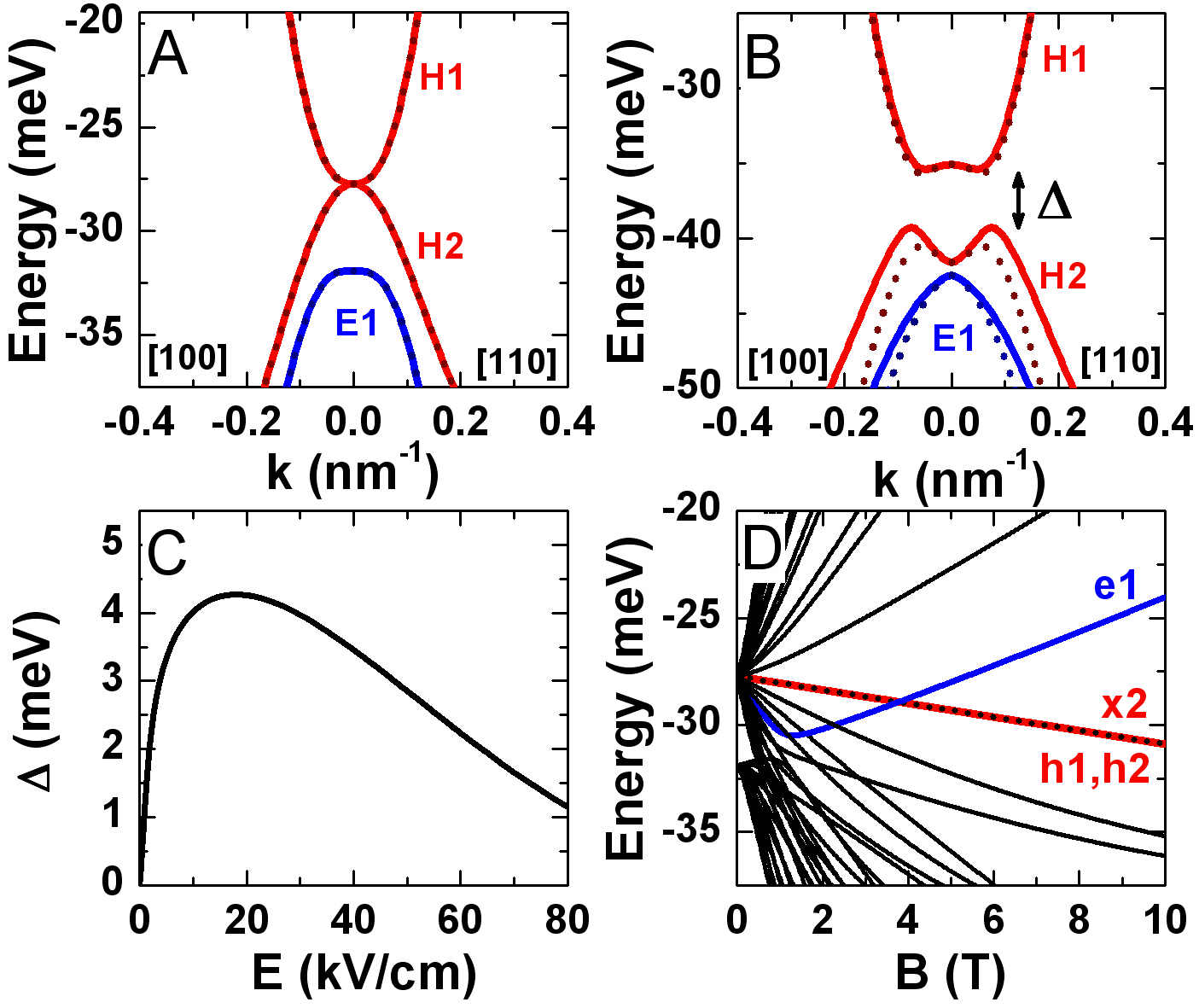} % Here is how to import EPS art
\caption{\label{Fig:3} (A,B) Band dispersions for BG phase, implemented in GaSb/InAs/GaSb QW at $d_1=10$~nm and $d_2= 6$~nm, in zero electric field (A) and in electric field of 5~kV/cm (B). Solid and dotted curves correspond to different spin states. (C) The band gap as a function of electric field. (D) The Landau-level fan chart. The zero-mode LL with doubled degeneracy order is marked by red bold curve (the \emph{h}1 and \emph{h}2 levels). The level \emph{e}1, containing only electron-like states in high magnetic fields from \emph{E}1 subband, is shown in blue.}
\end{figure}

\emph{BG phase in GaSb-designed QWs.}\textbf{--}The band structure in this phase mimics the structure of natural BG (see Fig.~\ref{Fig:3}A). In natural BG, a band gap can be opened by breaking of inversion symmetry between layers, for instance, by electric field perpendicular to the sample plane. The GaSb-designed QW also holds this property.

Fig.~\ref{Fig:3}B shows the band structure for the BG phase in electric field of 5~kV/cm. The dispersion curves are very similar to the band structure of natural BG in electric field~\cite{Q21}. However, the strong spin-orbit interaction in the GaSb-designed QWs removes the spin degeneracy at non-zero $k$ due to Rashba effect~\cite{Q23}. Fig.~\ref{Fig:3}C shows the band gap $\Delta$ as a function of electric field. It is seen that the band gap is electrically-tunable, as it is in natural BG, although its dependence on electric field in the GaSb-designed QWs has non-monotonic behaviour.

Another characteristic of natural BG is the unconventional quantum Hall effect~\cite{Q24}. For natural BG, plateaus in the Hall conductivity $\sigma_{xy}$, occur at integer multiples of $4e^2/h$, where the level degeneracy $g=4$ arises from the spin and valley degrees of freedom. Deviation from the conventional case occurs in the vicinity of the charge neutrality point, where there is a step in $\sigma_{xy}$ of height $8e^2/h$, arising from the eight-fold degeneracy of the zero-energy LL.

In the GaSb-designed QWs, a specific zero-mode LL with degeneracy two times higher than other LLs also arises. This zero-mode LL is presented in Fig.~\ref{Fig:3}D by the bold red curve (the \emph{h}1 and \emph{h}2 levels). The doubled degeneracy of such LL requires two times more carriers to fill it, so the transition between the corresponding plateaus should be twice as wide in density as others, and the $\sigma_{xy}$ step between the plateaus are expected to be $2e^2/h$ instead of $e^2/h$. As inversion asymmetry in the GaSb-designed QWs splits the zero-mode LLs, removing the double degeneracy order, we anticipate the recovery of the equidistant sequence of the plateaus such as for gate-biased natural BG~\cite{Q25}.

In the GaSb-designed QWs, there is another specific LL, which has a pure electron-like character at high magnetic field. The crossing of such level, denoted as \emph{e}1 in Fig.~\ref{Fig:3}D, with the zero mode LL, arising at critical magnetic field $B_c\simeq3.7$ T, corresponds to the transition from inverted into normal band ordering~\cite{Q7}.

The main difference between natural BG and the BG phase is that the electrons in the GaSb-designed QWs are not chiral particles although they mimic some characteristics of natural BG. Moreover, one can show~\cite{Q20} that the BG phase is a straight realization of the recently predicted state of matter, in which the gapless bulk states and spin-polarized edge channels coexist~\cite{Q26}. Previously, the BG phase was also predicted for two tunnel-coupled HgTe QWs~\cite{Q21}.

\begin{figure}
\includegraphics [width=1.0\columnwidth, keepaspectratio] {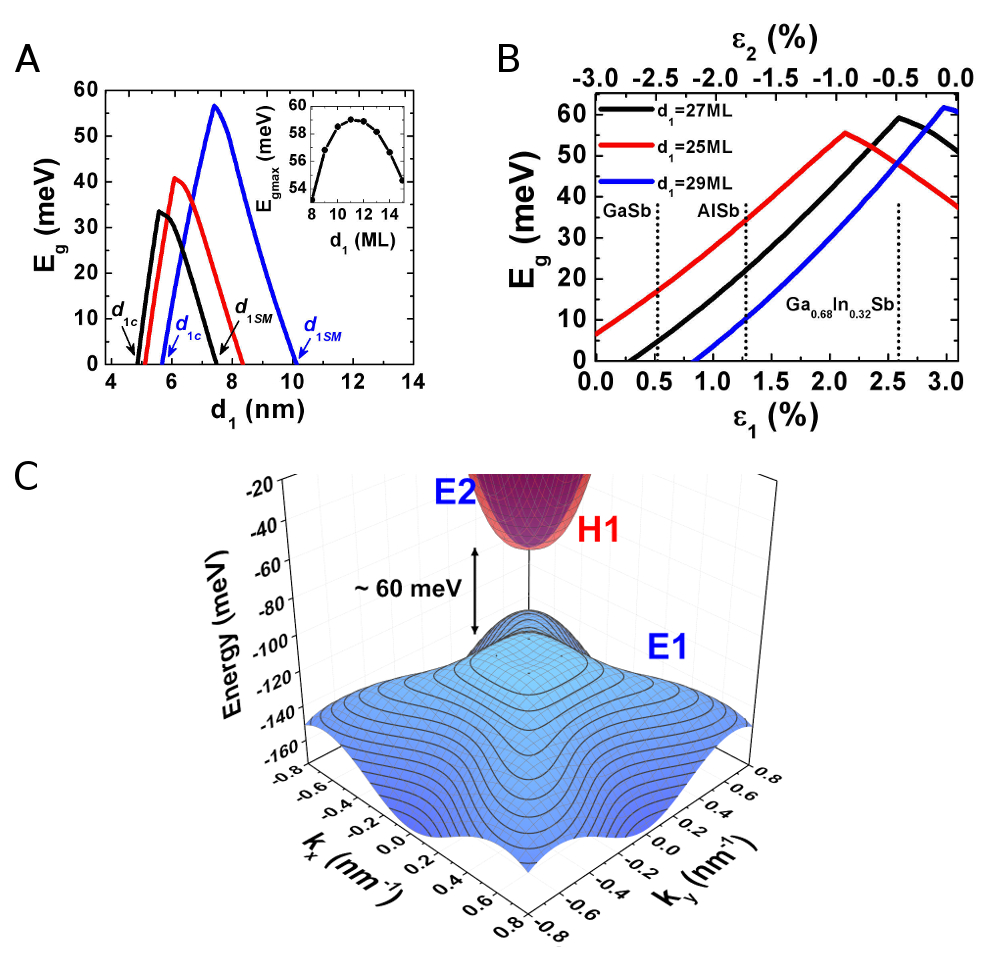} % Here is how to import EPS art
\caption{\label{Fig:4} (A) Inverted band gap in the InAs/Ga$_{0.6}$In$_{0.4}$Sb QW with InAs-design as a function of $d_1$ at $d_2=14$~ML. The black, red and blue curves correspond to the structures grown on GaSb, AlSb and Ga$_{0.68}$In$_{0.32}$Sb, respectively. The inset shows the maximum gap at given value of $d_2$ in the QW grown on Ga$_{0.68}$In$_{0.32}$Sb buffer. (B) Inverted band gap for the InAs/Ga$_{0.6}$In$_{0.4}$Sb QWs with different values of $d_1$ at $d_2=11$ ML as a function of $\varepsilon_1$ and $\varepsilon_2$. (C) A 3D plot of the band dispersion for the the QW with $d_1 = 27$~ML and $d_2 = 11$~ML grown on Ga$_{0.68}$In$_{0.32}$Sb buffer. The $x$ and $y$ axes are oriented along (100) and (010) crystallographic directions, respectively.}
\end{figure}

\emph{Large gap QSHI in InAs-designed QWs.}\textbf{--}We have demonstrated that InAs/GaSb/InAs and HgTe/CdTe QWs share the same phases. Now, we address the question if the inverted band gap in the three-layer QWs can be enhanced up to 55 meV, known for the compressively strained HgTe QWs~\cite{Q12}. To increase the band gap, we replace the middle GaSb layer by a GaInSb alloy~\cite{Q27}. Recently, this idea has been applied to the InAs/GaInSb QW bilayers~\cite{Q14,Q15}, in which the band gap is increased up to 20 meV for realistic indium content.

%Previously, Smith and Mailhiot found that such replacement enhanced the inverted band gap in strained InAs/GaInSb superlattice (SL)~\cite{Q27}.

Figure~\ref{Fig:4}A shows the inverted band gap in the three-layer InAs/Ga$_{0.6}$In$_{0.4}$Sb QW grown on various buffers as a function of $d_1$ at $d_2=14$~ML. The inset shows the maximum gap at given value of $d_2$ in the QW grown on Ga$_{0.68}$In$_{0.32}$Sb buffer. Figure 4B provides the inverted band gap for the InAs/Ga$_{0.6}$In$_{0.4}$Sb QWs with different values of $d_1$ and $d_2=11$~ML as a function of strain in InAs layers $\varepsilon_1$ and Ga$_{0.6}$In$_{0.4}$Sb layer $\varepsilon_1$. Figure~\ref{Fig:4}B demonstrates that by adjusting the strain, $d_1$ and $d_2$, one can obtain a QSHI with a band gap up to 60 meV. Indeed, the InAs/Ga$_{0.6}$In$_{0.4}$Sb QW with $d_1=27$~ML and $d_2=11$~ ML grown on Ga$_{0.68}$In$_{0.32}$Sb buffer (see Fig.~\ref{Fig:4}C) features the inverted band gap of about 60~meV. We note that the maximum band gap in QSHI state is achieved only symmetrical InAs/GaInSb/InAs QWs~\cite{Q20}.

Recent progress in the MBE technology of InAs/GaInSb structures~\cite{Q28,Q29} allows pseudomorphic growth of 40-period strained InAs/Ga$_{0.6}$In$_{0.4}$Sb SLs with good structural and surface quality. As InAs/Ga$_{1-x}$In$_x$Sb QWs do not require so large critical layer thickness, the pseudomorphic growth of InAs/Ga$_{1-x}$In$_x$Sb QW structures may be possible even at $x>0.4$. Therefore, an inverted band gap of more than 60 meV is expected in these QWs.

Finally, we stress an important advantage of three-layer InAs/GaInSb QWs as compared with HgTe/CdTe QWs. The inverted band gap in HgTe/CdTe QWs dramatically decreases with temperature~\cite{Q29b,Q30,Q31}. Therefore, a QSHI with the large gap in compressively strained HgTe/CdHgTe QWs is available at low temperatures only. Since temperature does not affect the ordering of the $\Gamma_6$ and $\Gamma_8$ bands in bulk InSb, InAs, GaSb semiconductors, we expect a weaker temperature effect on the band gap in the inverted InAs/GaInSb QWs.

\begin{acknowledgments}
~\\~ This work was supported by the Occitanie region via the "Gepeto Terahertz platform" and the ARPE project "Terasens", by the CNRS through the Emergence project 2016, and through the LIA "TeraMIR" project, by Montpellier University through the "Terahertz Occitanie Platform (TOP)" and by Russian Ministry of Education and Science (MK-1136.2017.2). Theoretical studies of the edge states were performed in the framework of project 16-12-10317 provided by Russian Science Foundation. Investigation of in multi-layered InAs/GaInSb QWs was supported by project 17-72-10158 from Russian Science Foundation.
\end{acknowledgments}

%\newpage
%\bibliography{InAs_GaSb_ref}

%

\end{document}